\documentclass{aa}
\usepackage{graphicx,color,subfigure,multirow}
\usepackage{txfonts}
\usepackage{verbatim}
\graphicspath{{./figures/}}
\usepackage{natbib,twoopt}
\usepackage[breaklinks=true]{hyperref}

\makeatletter

\newcommand{\Rmnum}[1]{\expandafter\@slowromancap\romannumeral #1@}
\makeatother

\hypersetup{
        colorlinks=true, 
        urlcolor=blue,
        linkcolor=blue
}
\bibpunct{(}{)}{;}{a}{}{,}

\makeatletter
\newcommand{\bibnote}[2]{\global\@namedef{#1note}{#2}}
\newcommand{\biblink}[2]{\global\@namedef{#1link}{#2}}

\makeatother
\makeatletter
\protected\def\stonyslink{%
        \def\hyper@linkstart##1##2{}\let\hyper@linkend\@empty}
\newcommandtwoopt{\citeads}[3][][]{%
        \href{http://adsabs.harvard.edu/abs/#3}%
        {\stonyslink \citealp[#1][#2]{#3}}%
        \biblink{#3}{\href{http://adsabs.harvard.edu/abs/#3}{ADS}}}
\newcommandtwoopt{\citepads}[3][][]{%
        \href{http://adsabs.harvard.edu/abs/#3}%
        {\stonyslink \citep[#1][#2]{#3}}%
        \biblink{#3}{\href{http://adsabs.harvard.edu/abs/#3}{ADS}}}
\newcommandtwoopt{\citetads}[3][][]{%
        \href{http://adsabs.harvard.edu/abs/#3}%
        {\stonyslink \citet[#1][#2]{#3}}%
        \biblink{#3}{\href{http://adsabs.harvard.edu/abs/#3}{ADS}}}
\newcommandtwoopt{\citeyearads}[3][][]{%
        \href{http://adsabs.harvard.edu/abs/#3}%
        {\stonyslink \citeyear[#1][#2]{#3}}%
        \biblink{#3}{\href{http://adsabs.harvard.edu/abs/#3}{ADS}}}
\makeatother

\title{Numerical estimation of the capture ability of Neptunian mean motion resonances}
\subtitle{\Rmnum{2}. Influence of eccentricity and inclination}

\author{
    Hailiang Li\inst{1} \email{lihailiang@must.edu.mo} \and   
    Li-Yong Zhou\inst{2}\fnmsep\inst{3} \email{zhouly@nju.edu.cn} \and  
    Xiaoping Zhang\inst{1}\fnmsep\inst{4} \corrauth{xpzhang@must.edu.mo}
        }
        
\institute{State Key Laboratory of Lunar and Planetary Sciences, Macau University of Science and Technology, Macau 999078, China
            \and
            School of Astronomy and Space Science, Nanjing University, 163 Xianlin Avenue, Nanjing 210046, China
              \and
            Key Laboratory of Modern Astronomy and Astrophysics in Ministry of Education, Nanjing University, China
              \and
            Macau University of Science and Technology Zhuhai MUST Science and Technology Research Institute, China
        }
\date{}
\titlerunning{aa61981-26}  
\authorrunning{Li et al.}
\begin{document}

\abstract{ 
Evaluating the ability of Neptunian mean-motion resonances (MMRs) to capture small bodies is an important step in reconstructing the evolutionary history of the early Solar System. Building on a previous planar model, this study extends the investigation of MMR capture during Neptune's migration to the three-dimensional case, systematically analyzing the effects of the small objects' eccentricity, $e$, and inclination, $i$. Through numerical simulations with different migration rates, we obtain the distribution of capture probabilities for 12 MMRs on the $e-i$ plane. The results show that increasing $i$ narrows the ``inevitable capture zone'' (strong MMRs and/or slow migration) and raises the lower $e$ threshold for capture (weak MMRs and/or fast migration), while it may also expand the upper $e$ limit for capture. When the resonance order reaches three, the $e-i$ plane further differentiates into several regions, which are closely related to the types of resonant angles. The emerging $i$-dominated region, along with its inevitable capture zone, makes third-order MMRs the most likely to benefit from higher inclination, as well as the most likely to host mixed-type resonators. Finally, using a hypothetical planetesimal distribution, we calculated the capture probability and efficiency for each MMR, finding that a cold disk favors strong MMRs, whereas a hot disk reduces differences among MMRs and across migration rates. In the future, a better determination of the population associated with different MMRs and different types of objects will help us understand the dynamical excitation state of the primordial disk in the early Solar System.

}

\keywords{celestial mechanics -- Kuiper belt: general -- methods: numerical
}
\maketitle{}
\nolinenumbers

\section{Introduction}
Trans-Neptunian objects (TNOs) generally refer to objects whose semimajor axes lie beyond that of Neptune. Due to their great distances and long dynamical timescales, they record and preserve a wealth of valuable history of the early Solar System. Among them, the resonant TNOs refer to those TNOs that reside in mean motion resonances (MMRs) with Neptune. \citetads{Smirnov2025} has pointed out that the trans-Neptunian region is profoundly influenced by MMRs, and potentially more than half of all TNOs reside in MMRs. In this paper, we use $p$:$q$ to denote the period ratio between Neptune and a resonant object, where $p<q$. Resonant TNOs are generally believed to have been captured into MMR during the planetary migration \citepads[e.g.,][]{Malhotra1995, Levison2008, Morbidelli2020}. The number and distribution of resonant TNOs reflect the dynamical evolution during the early planetary migration, as well as the stability over the subsequent billions of years.

A number of previous studies have used resonant TNOs as a diagnostic to assess the viability of planetary migration models. For instance, by adjusting parameters in planetary migration models, researchers have attempted to reproduce the eccentricity and inclination distributions of resonant TNOs \citepads[e.g.,][]{Malhotra1995, Gomes2000}, the existence of high-perihelion resonant objects \citepads[][]{Kaib2016}, and the leading-trailing asymmetry in the 1:$q$ MMRs \citepads[e.g.,][]{MurrayClay2005, Li2014b, Chen2019, Li2023}. On the other hand, the population of resonant TNOs is also a valuable diagnostic, reflecting both the capture efficiency of different MMRs \citepads[e.g.,][]{Chiang2002, Zhou2002, Nesvorny2016, Hermosillo2024} and the long-term stability of these MMRs over billions of years \citepads[e.g.,][]{Melita2000, Kotoulas2004, Tiscareno2009, Saillenfest2017, Li2024}.

With the accumulation of observational data, dozens of MMRs have now been identified to host TNOs. Some studies have provided estimates of the population of certain MMRs \citepads[e.g.,][]{Gladman2012, Adams2014, Volk2016, Crompvoets2022}. However, due to limited observational samples, the estimated populations for most MMRs still carry large uncertainties. Except for a few populous MMRs, there is little prospect of obtaining very precise observational estimates for most MMRs in the near future. Therefore, more detailed theoretical studies of the capture capability and stability of MMRs can help better interpret observational phenomena and predict the population of MMRs.

Since the understanding of resonance capture has become increasingly refined, studies based on analytical or semi-analytical methods have gradually expanded from adiabatic evolution \citepads[e.g.,][]{Henrard1982, Borderies1984} to nonadiabatic evolution \citepads[][]{Wyatt2003, Quillen2006, Mustill2011}, with most investigations focusing on low-order MMRs in the planar model. \citetads{Mustill2011} showed that for particles with low eccentricity, slow migration leads to inevitable capture, whereas fast migration leads to inevitable non-capture; for particles with a high eccentricity, capture occurs with a certain probability. Meanwhile, there are some preliminary conclusions about the effect of inclination on capture for a few specific MMRs \citepads[e.g.,][]{Li2014a, Li2014b, Li2020}.

In a recent study, \citetads{Li2026} used the eccentricity threshold for capture and the capture efficiency as indicators to measure the capture capability of a specific MMR. Given a migration rate and eccentricity distribution, this approach effectively enables comparison of capture numbers among different MMRs. However, the work of \citetads{Li2026} does not account for the effects of orbital inclination. On the other hand, the influence of eccentricity is limited to determining whether a planetesimal lies within the appropriate range for capture, and this treatment is simplified compared with those in previous studies \citepads[e.g.,][]{Borderies1984, Quillen2006, Mustill2011}.

In this paper, building on previous work, we aim to further evaluate the capture ability of Neptunian MMRs in the three-dimensional context. This study focuses on several major MMRs and investigates how planetesimal eccentricity and inclination affect resonant capture. In Section 2, we describe our simulation methods and initial conditions. In Section 3, we present the capture probability distributions under different eccentricities and inclinations. In Section 4, we discuss the distribution of resonance types. In Section 5, we combine the capture probability distributions with the orbital distribution of small bodies to compute their integrated capture probabilities and efficiencies, and attempt a more detailed comparison with observations. Finally, we summarize the results and provide a discussion in Section 6.

\section{Methods}

In this paper, we select a simple model consisting of the Sun, Neptune and test particles. All of the numerical simulations are performed with the N-body integration code REBOUND \citepads{rebound}, employing the WHFast algorithm \citepads[][]{wh,reboundwhfast}. The primary goal of this work is to investigate the influence of small body eccentricity, $e$, and orbital inclination, $i$. To achieve better resolution in the $e-i$ space, we focus on several specific MMRs, which requires first determining which MMRs to study.

For a given $p$:$q$ MMR, $k=|q-p|$ is called the order of the MMR. As summarized by \citetads{Li2026}, although small bodies have been detected in dozens of MMRs, most MMRs host only a few objects. If we focus on the MMRs for which the relative errors of the population estimates are small, it is easy to see that most of these MMRs have lower $k$ and $p$. Therefore, we select 12 MMRs from those with $k\leq5$, including three first-order MMRs (1:2, 2:3, and 3:4), two second-order MMRs (1:3 and 3:5), three third-order MMRs (1:4, 2:5, and 4:7), two fourth-order MMRs (1:5 and 3:7) and two fifth-order MMRs (2:7 and 3:8). Table \ref{tab:MMR_ini} lists the basic information of the 12 MMRs selected in this work.

\begin{table}[!htb]
    \centering
    \caption{Neptunian MMRs considered in this study.}
    \label{tab:MMR_ini}
    \begin{tabular}{ccccc}
        \hline
        MMR & $\alpha$ & Population $(D\gtrsim100\,$km) & $e_{max}$ & $a_{ini}$\,(au)\\
        \hline
        3:4 & 1.21 & $800^{+1100}_{-600}$ & 0.15 & $29.5\alpha$ \\
        2:3 & 1.31 & $10000^{+3600}_{-3000}$ & 0.3 & $-$ \\
        3:5 & 1.41 & $5000^{+5200}_{-3000}$ & 0.3 & $29.3\alpha$\\
        4:7 & 1.45 & $3000^{+4000}_{-2000}$ & 0.3 & $29.4\alpha$\\
        1:2 & 1.59 & $4400^{+1500}_{-1100}$ & 0.45 & $-$\\
        3:7 & 1.76 & $3000^{+5000}_{-2300}$ & 0.45 & $-$\\
        2:5 & 1.84 & $6600^{+4100}_{-3000}$ & 0.45 & $-$\\
        3:8 & 1.92 & $2300^{+5000}_{-2000}$ & 0.45 & $-$\\
        1:3 & 2.08 & $17000^{+11000}_{-8000}$ & 0.6 & $-$\\
        2:7 & 2.31 & $2300^{+5400}_{-1900}$ & 0.6 & $-$\\
        1:4 & 2.52 & $13000^{+15000}_{-8000}$ & 0.6 & $-$\\
        1:5 & 2.92 & $11000^{+19000}_{-8000}$ & 0.75 & $-$\\
        \hline
    \end{tabular}
    \tablefoot{The second column gives the ratio between the nominal location of each MMR and Neptune’s semimajor. The third column provides observational estimates of the population: the results for the 3:4, 3:5, and 4:7 MMRs are from \citetads{Gladman2012}, the result for the 2:3 MMR is from \citetads{Volk2016}, the result for the 1:2 MMR is from \citetads{Chen2019}; and the results for the remaining seven MMRs are from \citetads{Crompvoets2022}; all estimates correspond to TNOs with diameter $D\gtrsim100\,$km. The fourth column gives the maximum initial eccentricity explored in the simulations. The fifth column lists the initial semimajor of the planetesimals in the simulations, the dash indicates the same standard value ($29.5\alpha$) as the first row.}
\end{table}

If a planetesimal has an $e$ so high that its perihelion approaches Neptune's orbit too closely before the MMR reaches its semimajor, the planetesimal will be scattered first and cannot be captured. Based on a planar model, \citetads{Li2026} empirically pointed out that the eccentricity upper limit is roughly $1-\frac{32}{30\alpha}$, where $\alpha=(\frac{q}{p})^{\frac{2}{3}}$ denotes the ratio between the nominal MMR location and Neptune’s semimajor. This is roughly equivalent to requiring that the minimum perihelion for capture be at a distance of $\sim$2\,au beyond Neptune. Therefore, choosing appropriate ranges of $e$ and $i$ is crucial for achieving better spatial resolution in the $e-i$ plane. In this work, we set different $e$ ranges for different MMRs so that the $e$ covers the range 0 to $1-\frac{32}{30\alpha}$; the selected maximum eccentricities, denoted as $e_{max}$, are listed in the fourth column of Table \ref{tab:MMR_ini}. Meanwhile, for all MMRs the $i$ range is $0^\circ$ to $60^\circ$, which well covers the possible $i$ range of TNOs. The orbital elements $\Omega$, $\omega$, and $M$ of the planetesimals are all randomly selected from $0^\circ$ to $360^\circ$.

When calculating the capture probability at a given point in the $e-i$ plane, we select an elliptical region in the vicinity of that point and check the particles within this ellipse. The horizontal semi-axis of the ellipse is $\frac{1}{30}e_{max}$, which ranges from 0.005 to 0.025 depending on the specific MMR, and the vertical semi-axis is $2^\circ$. When placing the particles, their actual maximum $e$ and $i$ may slightly exceed $e_{max}$ and $60^\circ$, respectively. Specifically, in the $e-i$ plane, we evenly divide the eccentricity interval $[0, \frac{31}{30}e_{max}]$ and the inclination interval $[0, 62^\circ]$ into 31 equally spaced bins, yielding a total of $31 \times 31 = 961$ grid cells, with 50 particles randomly placed in each cell, for a total of $961 \times 50 = 48{,}050$ particles. In this way, under ideal conditions, when calculating the capture probability, the selected ellipse typically contains $\sim$157 particles, ensuring reasonably good statistical significance.

Because we aim to study specific MMRs independently, an important issue is to avoid interference from adjacent MMRs, especially those further out. To this end, we adopt a simplified, short-range migration model in which Neptune starts from 29\,au and is driven by an additional force to migrate to 30\,au at a constant rate, $\dot{a}_N$, after which it stops. During this process, Neptune maintains a circular orbit. We consider four different values of $\dot{a}_N$: 0.01, 0.1, 1, and 10\,au/Myr. Although this migration model does not incorporate the damping characteristic of the traditional exponential migration model \citepads[e.g.,][]{Malhotra1993}, nor does it account for the dramatic orbital changes that may result from giant planet instability \citepads[e.g.,][]{Tsiganis2005, Nesvorny2012}, the range of $\dot{a}_N$ nevertheless covers the plausible values in actual scenarios, and each $\dot{a}_N$ can be regarded as corresponding to a particular stage of a ``realistic'' migration model.

Since $\dot{a}_N$ is a constant within each run, it is not necessary to consider an initial semimajor axis distribution for the planetesimals. By default, we set the initial semimajor of all particles, $a_{ini}$, to $29.5\alpha$\,au, which means that they can be captured when Neptune has migrated halfway (owing to the width of the MMR, the actual time is always slightly earlier). A captured particle will subsequently migrate outward with Neptune and eventually reach the nominal MMR location $30\alpha$\,au. After Neptune arrives at 30\,au and stops migrating, the system is then integrated for 0.5\,Myr with the extra force removed. It is worth mentioning that, although the removal of this extra force causes an abrupt change in $\dot{a}_N$, the resonant particles have already been evolving within the MMR for some time, so their resonant state is not altered by this change \citepads[see e.g.,][]{Li2023}. 

For each small body, we record its average semimajor $\overline{a}$ over the last 0.5\,Myr and use it to determine whether it has been captured. This identification method based on the $\overline{a}$ has been shown to be effective by \citetads{Li2026}. Specifically, we consider a particle to be captured if $|\overline{a}- 30\alpha| \leq 0.02$\,au, whereas we consider it non-captured if $\overline{a} < 30\alpha - 0.02$\,au. Clearly, besides these two categories, there is also a class of particles with $\overline{a} > 30\alpha + 0.02$\,au; these bodies have not been swept by the primary MMR and thus never have the opportunity to be captured. We refer to them as disturbed objects. In addition, a small number of particles are scattered out of the Solar System during the simulation. Most of the disturbed and scattered objects possess exceptionally high initial eccentricities.

The vicinity of any MMR is always densely filled with high-order MMRs; those whose locations lie between the main MMR and $a_{ini}$, i.e., between $29\alpha$\,au and $29.5\alpha$\,au, will sweep over the particles first and carry them away. According to the empirical rule obtained by \citetads{Li2026}, the minimum eccentricity threshold required for resonant capture ($e_{min}$) increases with $k$, $\alpha$, and $\dot{a}_N$. Therefore, the interference from these high-order MMRs tends to manifest in high-$e$ regions and is correspondingly reduced when $\dot{a}_N$ is faster. Typically, the 1:$q$ MMRs and those distant MMRs are relatively little affected. Taking the 1:2 MMR as an example, on average only 0.53\% of the particles end up with $\overline{a} > 30\alpha + 0.02$\,au across the four simulations. Upon inspection, the dominant interference comes from the 19:39 MMR, a 20th-order MMR. For the 3:4 MMR, however, 13.59\% of the particles are disturbed by exterior MMRs, with the main interference coming from the 14:19 MMR, a fifth-order MMR. When calculating the capture probability, we consider only the aforementioned captured and non-captured small bodies, thereby minimizing such interference. Nevertheless, such interference still leads to a certain degradation of the statistical significance of the capture probability calculation in the high-$e$ region. The influence of adjacent MMRs is also reflected in the fact that even if they do not capture particles, they will still slightly alter their orbits \citepads[e.g.,][]{Mustill2011}. After a thorough examination, we ignore this effect, because these adjacent MMRs all have very high orders.

Another more severe interference occurs when a higher-order MMR is located particularly close to $a_{ini}$: particles can be directly locked in this higher-order MMR. This phenomenon is especially pronounced for the two MMRs situated within the classical main belt of TNOs. Taking the 4:7 MMR as an example, if one simply sets $a_{ini} = 29.5\alpha \approx 42.84$\,au, then 27.87\% of the particles are disturbed and lose their chance of being captured (given that they predominantly occupy high-$e$ regions, this proportion is quite significant), with the vast majority of the interference coming from the 5:9 MMR (whose initial nominal location is 42.91\,au). To avoid this, when studying the 4:7 MMR, we slightly shifted the $a_{ini}$ inward to $29.4\alpha$\,au, thereby reducing the fraction of disturbed particles to 6.07\%. Similarly, when studying the 3:5 MMR, the $a_{ini}$ is reduced to $29.3\alpha$\,au to avoid the main interference from the 7:12 MMR. Meanwhile, we have carefully checked that the slight reduction in $a_{ini}$ does not place any object initially inside the 3:5 MMR or the 4:7 MMR.

In the three-dimensional case, an MMR can have multiple defined resonant angles. For any given MMR, its resonant angle can be written as $\phi = q\lambda - p\lambda_N - (q-p-2j)\varpi - 2j\Omega$, where $j$ denotes a nonnegative integer. Since the resonance strength is proportional to $e^{|q-p-2j|} i^{|2j|}$, it is usually sufficient to consider resonant angle forms that minimize $|q-p-2j| + |2j|$, which requires $2j \leq q-p$. Therefore, for first-order MMRs, $j$ can only take the value 0; for second- and third-order MMRs, $j$ can be 0 and 1; and for fourth- and fifth-order MMRs, $j$ can be 0, 1, or 2. To distinguish among different $\phi$, we denote the $\phi$ corresponding to a specific value of $j$ as $\phi_j$. Thus, $\phi_0 = q\lambda - p\lambda_N - (q-p)\varpi$ corresponds to a pure eccentricity-type MMR, whereas $\phi_1$ and $\phi_2$ correspond to inclination-type or mixed-type MMRs.

\section{Distribution of capture probabilities}

\subsection{First-order resonances}
\begin{figure}[!htb]
\centering
\resizebox{\hsize}{!}{\includegraphics{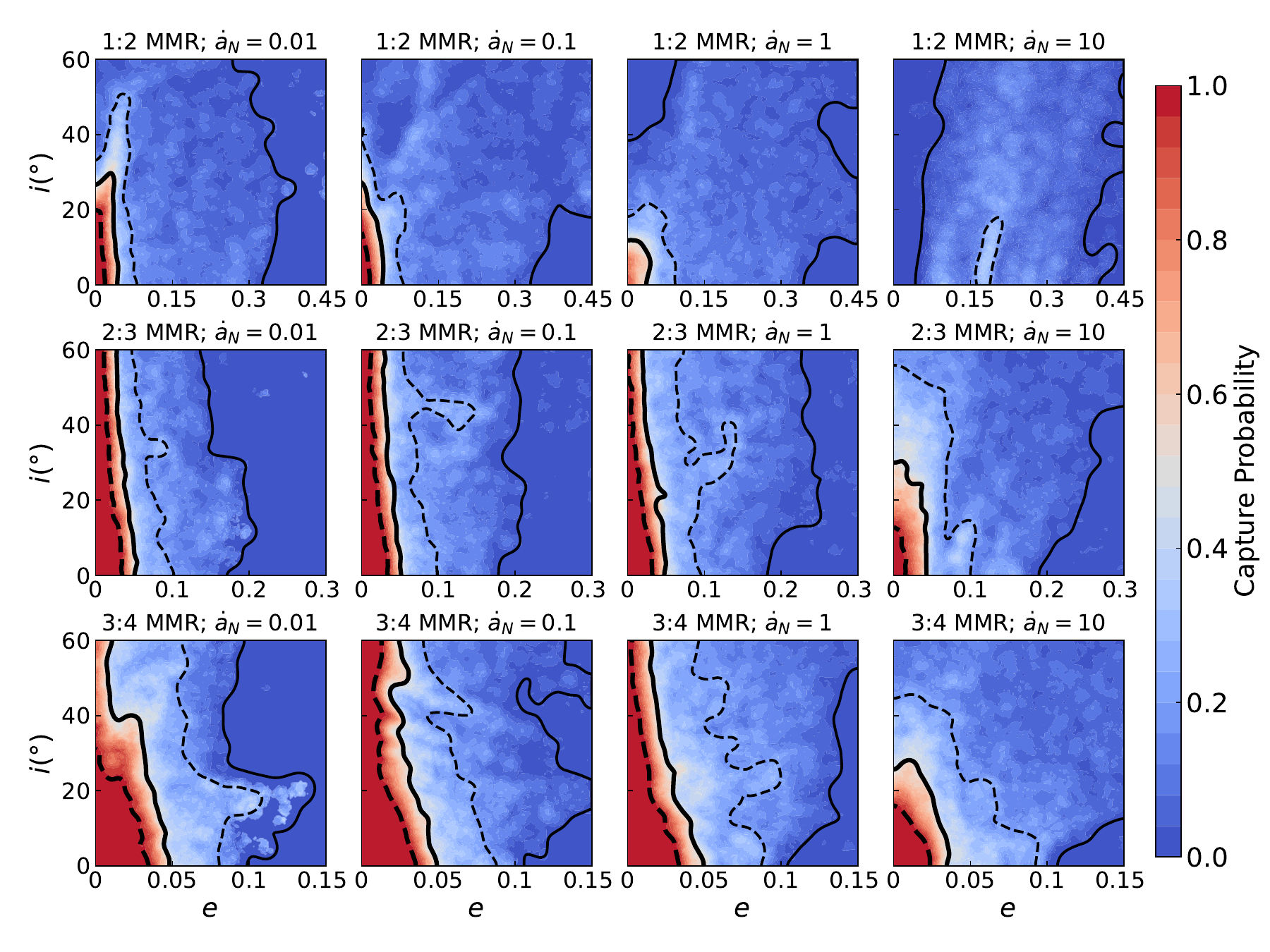}}
\caption{Capture probability distribution in the $e$–$i$ plane for the first-order MMRs. The horizontal and vertical axes represent eccentricity and orbital inclination, respectively, and the color indicates the capture probability. The thick dashed, thick solid, thin dashed, and thin solid lines represent the 97\%, 50\%, 20\%, and 2\% contours, respectively. The three different first-order MMRs are shown from top to bottom. The four different $\dot{a}_N$ are shown from left to right, with $\dot{a}_N$ in units of au/Myr.}
\label{fig:ei_dis_1st}
\end{figure}

In the previous section, we described how to exclude disturbed particles and how to distinguish between captured and non-captured ones. Using only the captured and non-captured particles, we calculate the capture probability at different locations. Since the distribution of capture probability in the $e-i$ plane depends primarily on $k$, we organize our results according to $k$. We first present the capture probability distribution for the first-order MMRs in Fig.~\ref{fig:ei_dis_1st}, where we have also marked the positions of the contour lines. A certain level of smoothing is applied when plotting these contour lines.

It has been known that, in the planar case, first-order MMRs can achieve a 100\% capture probability at low $e$ \citepads[e.g.,][]{Borderies1984}. Once the $e$ exceeds a certain threshold, the capture probability drops rapidly to below 50\%, as evidenced by the proximity of the contour lines corresponding to 97\% and 50\%. Beyond this point, the capture efficiency declines slowly with increasing $e$ and approaches zero when $e \gtrsim 1 - \frac{32}{30\alpha}$.

With the introduction of inclination, the $e$ range of inevitable capture zone gradually narrows, indicating that an increase in $i$ is unfavorable for capture. For the 1:2 MMR, when $\dot{a}_N = 0.01$\,au/Myr, the inevitable capture zone can extend up to $20^\circ$, but it completely disappears when $\dot{a}_N \geq 1$\,au/Myr. For the 2:3 and 3:4 MMRs, the inevitable capture zones still exist, though their area is somewhat reduced, and no longer cover the high-$i$ range when $\dot{a}_N = 10$\,au/Myr.

If we examine the contour line for 2\% capture probability, the influence of inclination appears to be intricate. When $\dot{a}_N = 10$\,au/Myr, the contour tilts toward the high-$e$ region as $i$ increases, indicating that MMRs can capture particles even when $e \gtrsim 1 - \frac{32}{30\alpha}$. This suggests that, in this regime, an increase in $i$ is favorable for capture. The failure of MMR capture at excessively high $e$ is primarily caused by Neptune's scattering effect, and a higher $i$ reduces the frequency of close encounters between the object and Neptune, thereby helping avoid scattering. A similar phenomenon has already been pointed out in previous work \citepads[e.g.,][]{Li2020}. For the slower cases ($\dot{a}_N = 1$ or 0.1\,au/Myr), the tilt of the 2\% contours is no longer evident. In the slowest case ($\dot{a}_N = 0.01$\,au/Myr), the capture probability first increases and then decreases with rising $i$.

For the 1:2 MMR, in the fastest migration case, the capture can no longer occur at $e = 0$ but instead requires a certain $e_{min}$. This phenomenon has been noted by \citetads{Li2026} and may be attributed to the special structure of the 1:$q$ MMRs or their large distances. It can also be observed that the value of $e_{min}$ slightly increases with $i$, meaning that in this case an increase in $i$ is again unfavorable for capture.

If we examine the entire $e-i$ plane across all values of $\dot{a}_N$, the integrated capture probabilities for the 1:2, 2:3, and 3:4 MMRs are found to be 9\%, 18\%, and 26\%, respectively. However, it should be noted that this does not imply that the 3:4 MMR possesses the strongest capture capability. In fact, the main reason is that we adopted different values of $e_{max}$ for these MMRs. Inspecting the horizontal axis scales reveals that the capture range of the 3:4 MMR extends to $e \approx 0.1$, whereas the 2:3 and 1:2 MMRs extend to about 0.2 and 0.3, respectively. Therefore, it is meaningful to compare their capture probability under a unified $e-i$ distribution of particles; this will be discussed in a later section of this paper.

\subsection{Second-order resonances}

\begin{figure}[!htb]
\centering
\resizebox{\hsize}{!}{\includegraphics{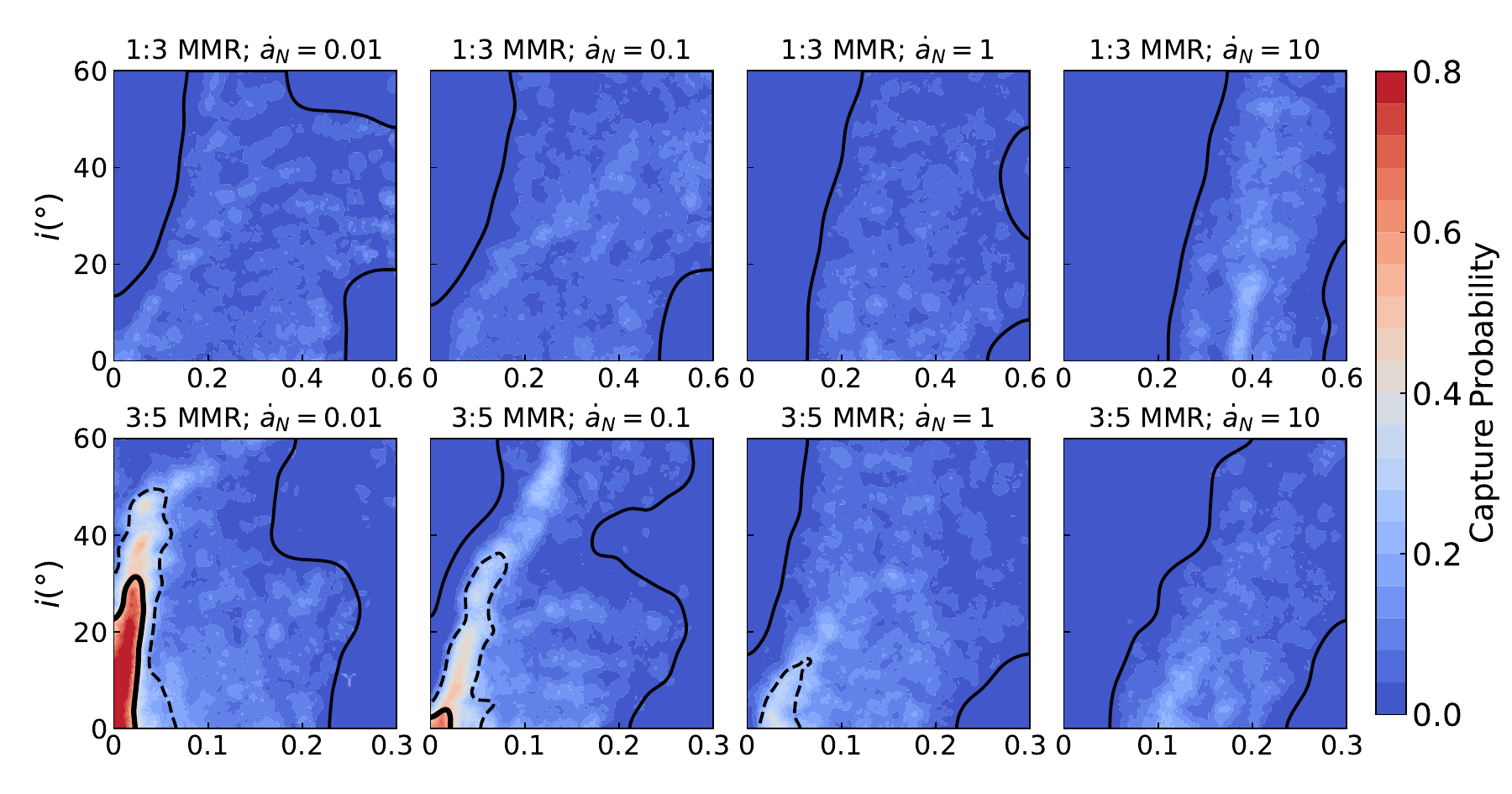}}
\caption{Same as Fig.~\ref{fig:ei_dis_1st} but for the second-order MMRs. Only the 50\%, 20\%, and 2\% contours are shown.}
\label{fig:ei_dis_2nd}
\end{figure}

The capture probability distributions of the second-order MMRs are shown in Fig.~\ref{fig:ei_dis_2nd}. As a class of MMRs with intermediate strength, the second-order MMRs exhibit capture capabilities in a transitional state. Among them, the capture probability distribution of the 3:5 MMR bears some resemblance to that of the first-order MMRs. For $\dot{a}_N = 0.01$\,au/Myr, the 3:5 MMR can achieve a capture probability exceeding 50\% in the relatively low-$e$ and low-$i$ region; moreover, as $e$ and $i$ approach 0, the capture probability can approach nearly 100\%. On the other hand, the 1:3 MMR is more similar to the higher-order MMRs; its capture probability remains generally below 10\% throughout the $e-i$ plane.

We have previously shown that the 1:2 MMR cannot capture particles at $e = 0$ when $\dot{a}_N = 10$\,au/Myr, but instead requires a certain $e_{min}$. For the second-order MMRs, we can observe more clearly how the eccentricity threshold is affected as $\dot{a}_N$ increases. For the 3:5 MMR, when migration is sufficiently slow, capture can occur at $e = 0$. When $\dot{a}_N = 0.01$\,au/Myr, this conclusion holds from $i = 0^\circ$ to beyond $60^\circ$. When $\dot{a}_N = 0.1$\,au/Myr or 1\,au/Myr, this holds only at low inclination. When $\dot{a}_N = 10$\,au/Myr, the 3:5 MMR can no longer capture particles with $e = 0$. The situation is similar for the 1:3 MMR: it can capture particles on circular orbits at low inclinations only when $\dot{a}_N = 0.01$\,au/Myr, but cannot do so when migration is fast.

For the 3:5 MMR, within the region suitable for capture, a higher capture probability can be observed at the low-$e$ region, and this phenomenon is especially pronounced in the slow migration cases ($\dot{a}_N \leq 0.1$\,au/Myr). For fast migration, the enhancement at low-$e$ becomes less distinct, and the capture probability becomes relatively low and similar across different values of $e$ and $i$. For the 1:3 MMR, this phenomenon is very weak: within the capture region enclosed by the dashed contours, the capture probability is nearly uniform.

\subsection{Third-order resonances}

\begin{figure}[!htb]
\centering
\resizebox{\hsize}{!}{\includegraphics{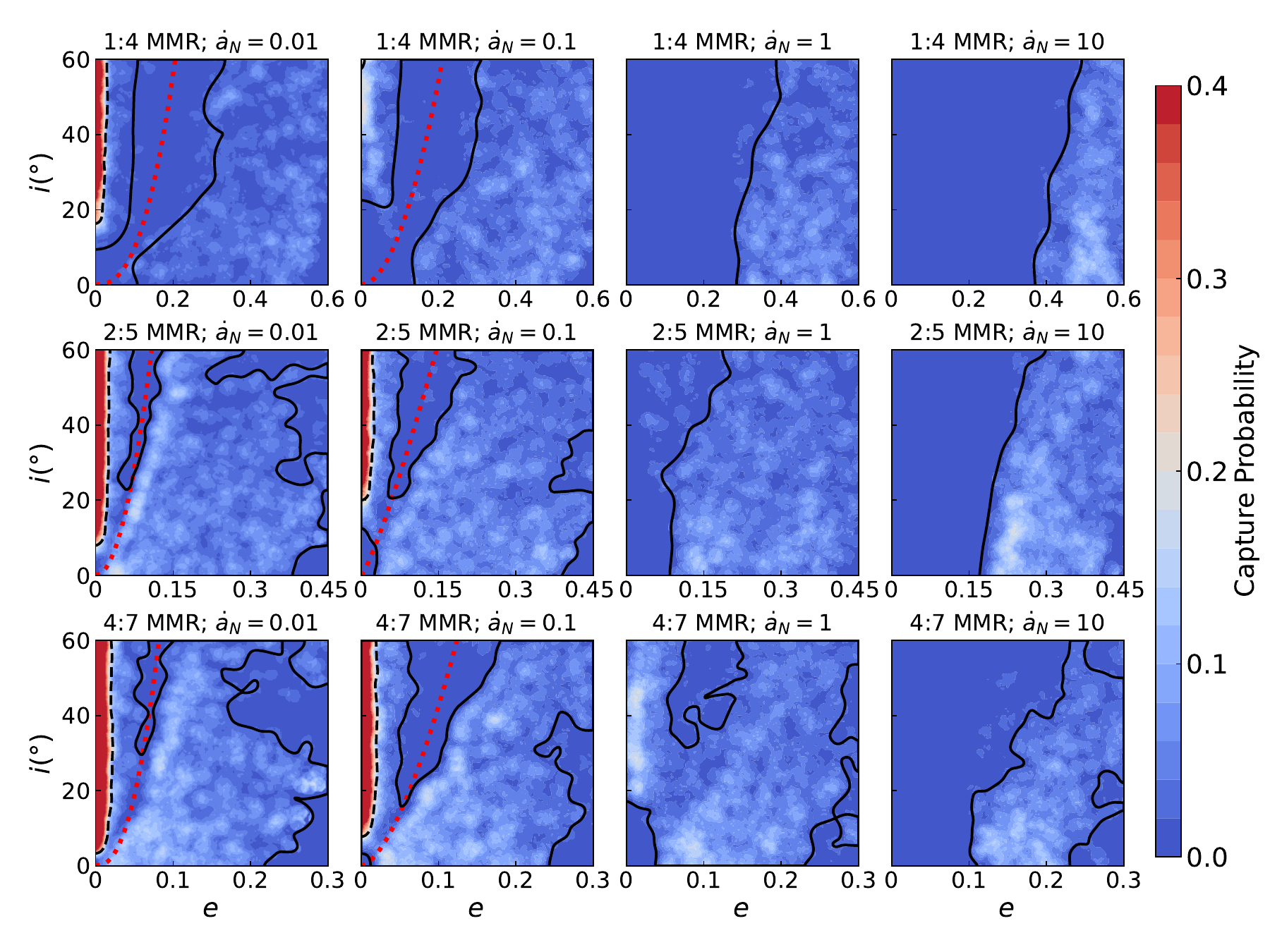}}
\caption{Same as Fig.~\ref{fig:ei_dis_1st} but for the third-order MMRs. Only the 20\% and 2\% contours are shown. The dotted red line roughly marks the boundary between the $e$-dominated region and the $i$-dominated region.}
\label{fig:ei_dis_3rd}
\end{figure}

The probability distributions for the third-order MMRs are shown in Fig.~\ref{fig:ei_dis_3rd}. A prominent feature of the third-order MMRs is that, at low $\dot{a}_N$, the capture region is divided into two major parts: one located in the upper-left region of the $e-i$ plane, and the other in the lower-right region. Between these two regions, there is also a ``gap'' where the capture probability is very low. In this paper, we refer to the lower-right capture region as the ``$e$-dominated region'' and the upper-left region as the ``$i$-dominated region.'' These two regions correspond largely to the two types of resonant angles, which we will discuss in detail in the next section. In Fig.~\ref{fig:ei_dis_3rd}, the dotted red line roughly marks the boundary between the two regions. This boundary is obtained through a fitting procedure, using a simple power-law relationship.

\begin{figure}[!htb]
\centering
\resizebox{\hsize}{!}{\includegraphics{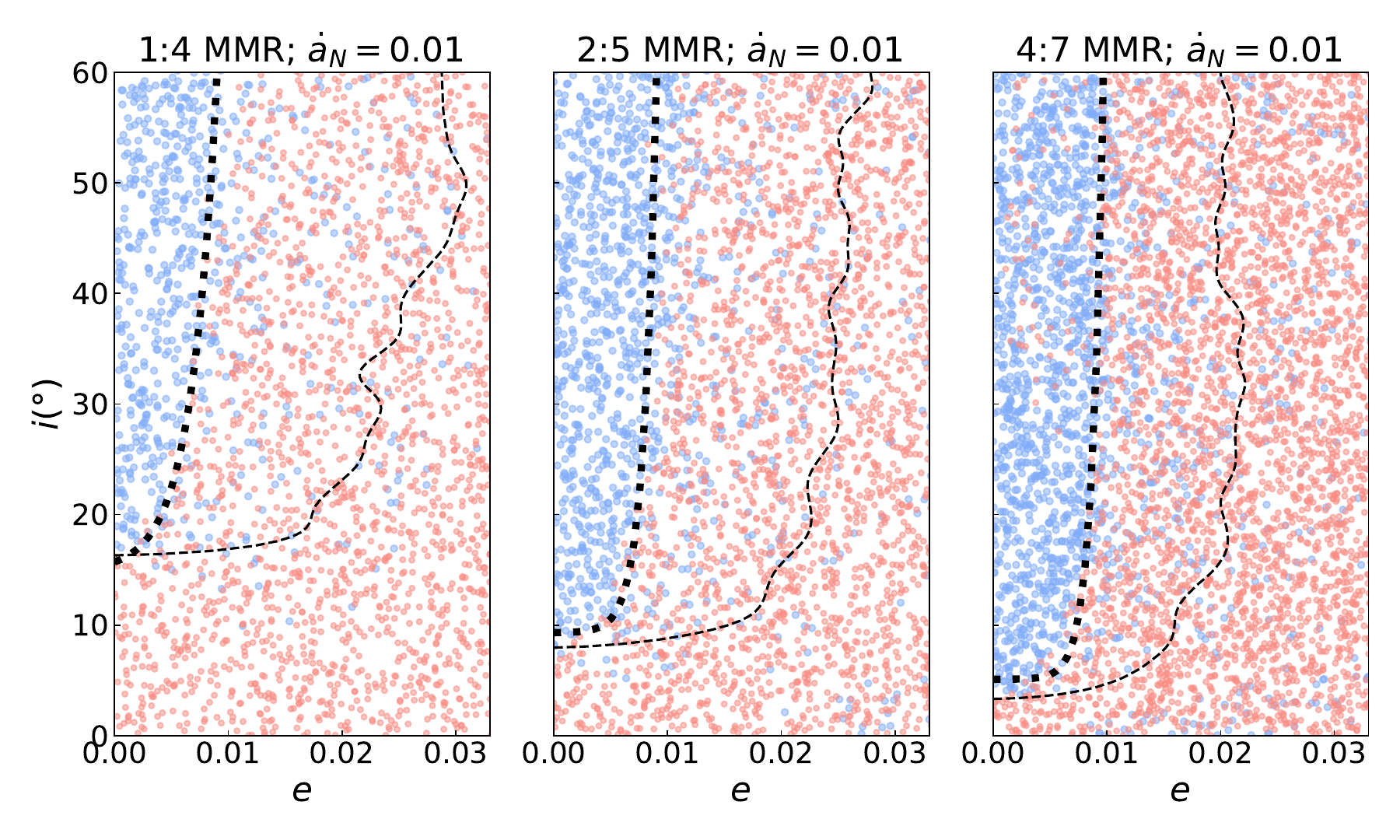}}
\caption{Capture outcomes in the low-$e$ region for several third-order MMRs when $\dot{a}_N = 0.01$\,au/Myr. The blue and red points represent captured and non-captured particles, respectively. The thick dotted line indicates the approximate boundary within which small bodies are inevitably captured, and the thin dashed line represents the 20\% capture probability contour.}
\label{fig:boundfind}
\end{figure}

At the leftmost edge of the $i$-dominated region (i.e., at the lowest $e$), the capture probability is extremely high. Fig.~\ref{fig:ei_dis_3rd} only shows the 20\% capture probability contour, but in fact the capture probability in this region can reach 100\%. To illustrate this, we show in Fig.~\ref{fig:boundfind} the capture outcomes for particles in this narrow region when $\dot{a}_N = 0.01$\,au/Myr. It can be seen that when $e<0.01$ and $i$ exceeds a certain value, the third-order MMRs can capture 100\% of planetesimals. We use a simple exponential function to fit the approximate boundary of the ``inevitable capture region'', which is shown as the dashed line in Fig.~\ref{fig:boundfind}. It can be seen that on the upper-left side of the dashed line, almost no particles escape capture.

This inevitable capture region is very narrow. In Fig.~\ref{fig:boundfind} we also show the same 20\% capture probability contour as in Fig.~\ref{fig:ei_dis_3rd}. As the $e$ increases slightly, the capture probability drops rapidly to below 20\%, and when $e>0.03$, the capture probability decreases further to only about 5\%. Furthermore, when $i\gtrsim30^\circ$, the 4:7 MMR can no longer guarantee capture for objects in this region—there are a few exceptions—but the capture probability still exceeds 90\%, and a clear difference in capture probability can be seen on the two sides of the dashed line.

By comparing on the same scale, we can also observe that the area of inevitable capture for the three third-order MMRs expands slightly with decreasing $\alpha$ and gradually extends toward the low-$i$ region, with the 4:7 MMR having the widest extent. In terms of $e$, however, the areas of inevitable capture are comparable, with the $e$ never exceeding 0.01. From Fig.~\ref{fig:ei_dis_3rd}, it is also evident that even the entire $i$-dominated region is relatively narrow compared with the $e$-dominated region.

As the $\dot{a}_N$ increases, the inevitable capture region disappears first; at this stage, the $i$-dominated region shifts somewhat toward higher $e$, as shown in Fig.~\ref{fig:ei_dis_3rd} for the 1:4 MMR at $\dot{a}_N = 0.1$\,au/Myr or for the 4:7 MMR at $\dot{a}_N = 1$\,au/Myr. When $\dot{a}_N$ increases further, the $i$-dominated region eventually vanishes completely, and the capture probability distributions of the third-order MMRs become similar to those of the second-order MMRs.

Previous studies have reached relevant conclusions for low-order MMRs in the planar case, indicating that inevitable capture occurs under low $e$ and slow migration \citepads[e.g.,][]{Borderies1984, Mustill2011}. The inevitable capture region of the third-order MMRs follows the same basic condition, but with additional complexity. In the planar case, low $e$ implies a weaker resonance strength, yet here, when both $e$ and $i$ approach zero, capture does not occur, which is completely different from the inevitable capture seen in the 3:5 MMR or in the first-order MMRs. In particular, the inevitable capture for third-order MMRs requires extremely low $e$ and a certain $i$ simultaneously; therefore, one cannot attribute the phenomenon to a ``strong'' or ``weak'' resonance strength. Moreover, considering the form of the resonant angle, the second-order MMRs also possess a corresponding inclination-type MMR with the resonant angle $\phi_1 = (p+2)\lambda - p\lambda_N - 2\Omega$, whereas for the third-order MMRs, $\phi_1 = (p+3)\lambda - p\lambda_N - \varpi - 2\Omega$ corresponds to a mixed-type MMR. Why it is the third-order MMRs that exhibit a separation into $e$-dominated and $i$-dominated regions remains a question that deserves further study.

Overall, although the $i$-dominated region is very narrow, the presence of the inevitable capture region allows it to achieve a very high capture probability. When $\dot{a}_N = 0.01$\,au/Myr, if all captured particles are divided into two parts according to the red dashed line in Fig.~\ref{fig:ei_dis_3rd}, then for the 1:4, 2:5, and 4:7 MMRs, the particles to the left of the dashed line account for 48.4\%, 45.6\%, and 53.5\% of the total, respectively. These fractions, however, are obtained under the premise that we are currently discussing a uniform $e$ and $i$ distribution. Whether the exceptionally high capture probability of $i$-dominated capture would significantly affect the captured populations of the third-order MMRs in a realistic planetesimal disk will be discussed later in this paper.

\subsection{Higher-order resonances}

\begin{figure}[!htb]
\centering
\resizebox{\hsize}{!}{\includegraphics{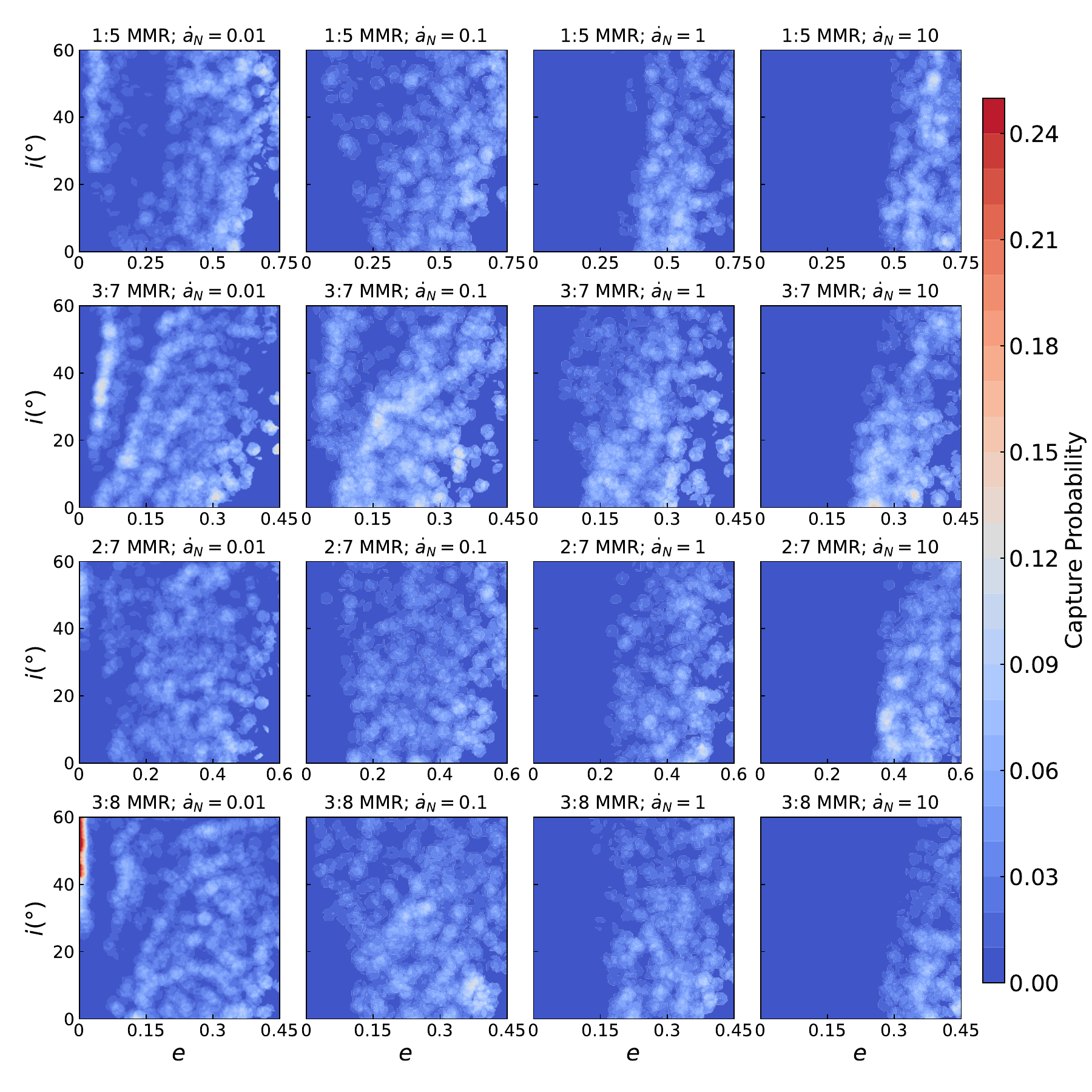}}
\caption{Same as Fig.~\ref{fig:ei_dis_1st} but for the higher-order $k\geq4$ MMRs. No contours are shown in this figure.}
\label{fig:ei_dis_high}
\end{figure}

Finally, we present the capture probability distributions for several higher-order MMRs, as shown in Fig.~\ref{fig:ei_dis_high}. These MMRs have only low capture probabilities, generally below 10\%, which makes their distributions appear rather fragmented.

Among them, the two fourth-order MMRs are similar to the third-order MMRs: when migration is slow, the entire $e-i$ plane can be divided into $e$-dominated and $i$-dominated regions. However, the left edge of the $i$-dominated capture region is not in close contact with $e = 0$ as in the third-order MMRs. Instead, they are slightly offset from $e = 0$, similar to the fast-migration cases of the third-order MMRs (e.g., the 4:7 MMR at $\dot{a}_N = 1$\,au/Myr). This may be because $\dot{a}_N = 0.01$\,au/Myr is still relatively fast for these two fourth-order MMRs.

For the fifth-order MMRs, the most striking feature appears in the case of $\dot{a}_N = 0.01$\,au/Myr, where the additional differentiation occurs in the upper-left corner of the $e-i$ plane. Specifically, the upper-leftmost region is again an $i$-dominated region, followed by a gap of low capture probability. To the lower-right of this gap lies another region with slightly higher capture efficiency. To the lower-right of this intermediate region is yet another gap, followed by the $e$-dominated region that occupies the remaining area.

It should be noted that the two fifth-order MMRs are both more distant than the 3:7 MMR and higher in order, yet their $i$-dominated region can appear at $e \approx 0$. To be precise, the newly emerging capture region is the one at the very top-left corner, rather than the one sandwiched in between. Therefore, the above terminology may be somewhat ambiguous, because the ``$i$-dominated region'' here is not the same region defined in the third- and fourth-order MMRs. Perhaps a better naming scheme would be to use the parameter $j$ that is employed in defining the $\phi_j$. For the closer 3:8 MMR, within the newly formed capture region at the top-left corner, there is also a narrower zone ($i \gtrsim 40^\circ, e \lesssim 0.002$) where the 3:8 MMR possesses the ability to inevitably capture objects. This is broadly similar to the inevitable capture region of the third-order MMRs at low $e$ and high $i$, except that its range is much narrower.

The differentiation phenomenon observed in the fifth-order MMRs is very similar to that of the third-order MMRs, and it naturally leads to another similar phenomenon: as the $k$ increases, the number of possible $j$ values also increases. The differentiation of capture regions does not occur first in even-order MMRs, which implies that pure inclination-type MMRs do not independently form a separate capture region, whereas mixed-type MMRs can. Based on this rule, we conjecture that when $k$ reaches $2n+1$, a total of $n+1$ capture regions will form in the $e-i$ plane. In the following sections, we will further explore how the different capture regions relate to the resonant angle of the particles.

\section{Distribution of resonance types}

So far, we have presented the capture probability distributions for different MMRs at various $\dot{a}_N$ across the $e-i$ plane. Next, we will further discuss the distribution of different resonance types. Here, the term ``resonance type'' refers to the type of the resonant angle form, which classifies resonances into eccentricity-type, inclination-type, or mixed-type resonances.

In Section 2, we used different values of $j$ to define the $\phi_j$. Generally, for a given MMR, the most common form is the eccentricity-type MMR, with $\phi_0 = q\lambda - p\lambda_N - (q-p)\varpi$. We therefore adopt distinct criteria when determining whether a given $\phi_j$ is librating. For the pure eccentricity-type MMR, $\phi_0$ is considered to be librating if its libration amplitude $\Delta\phi_0 < 350^\circ$. For inclination-type and mixed-type MMRs, i.e., when $j \ge 1$, we require $\Delta\phi_j < 300^\circ$. Because the libration of $\phi_1$ is sometimes caused jointly by the $\phi_0$ libration and the slow precession of the $\omega$, this differentiated criterion helps to exclude some ``spurious'' libration. In fact, genuine $\phi_1$ libration usually has a lower amplitude, because most of these particles reside in the low-$e$ region.

Based on the libration amplitude criteria, we classify the particles into several types: if only one $\phi_j$ is librating, we refer to such a particle as a $\phi_j$ resonator, and if multiple resonant angles are librating simultaneously, we call them $\phi_{multi}$ resonators. It should be noted that the classification here only reflects the properties of the particles at the early stage of capture. Whether different types of resonators can survive stably to the present day in the complex dynamical environment of the Solar System requires further investigation, and the possibility that resonators may switch between types cannot be ruled out.

\begin{figure*}[!htb]
\centering
\resizebox{\hsize}{!}{\includegraphics{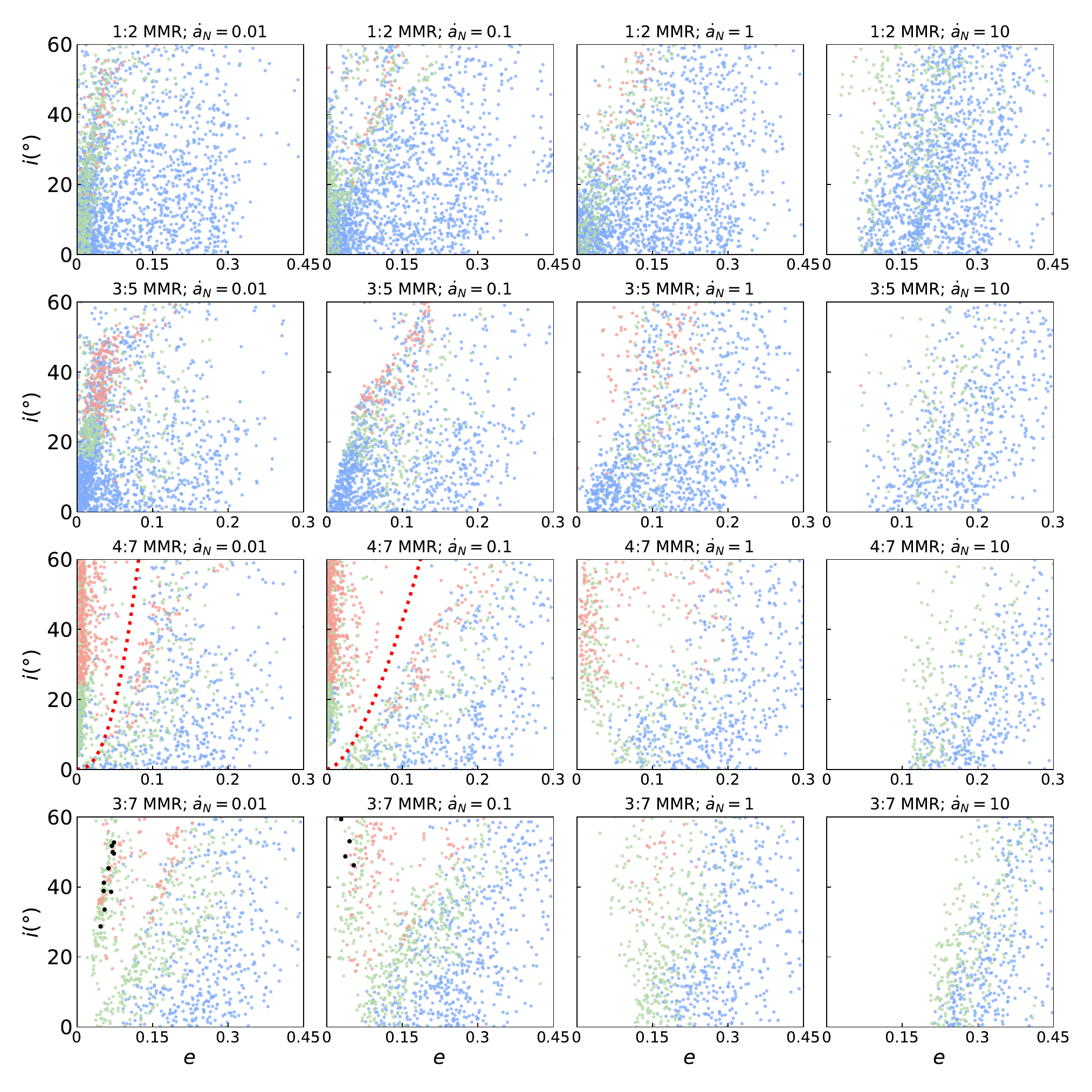}}
\caption{Distributions of resonant angle types for particles captured into the 1:2, 3:5, 4:7, and 3:7 MMRs. Blue, red, and black dots (only appear in the 3:7 MMR) represent objects with librating resonant angles $\phi_0$, $\phi_1$, and $\phi_2$, respectively, while green points indicate objects with multiple resonant angles librating simultaneously. For clarity, in the 1:2, 3:5, and 4:7 MMRs, only half of the particles are randomly selected for plotting. For the 4:7 MMR in the slow migration case, the dotted red line marks the approximate boundary between the $e$-dominated and $i$-dominated regions.}
\label{fig:type_phi}
\end{figure*}

We select four representative MMRs—the 1:2, 3:5, 4:7, and 3:7 MMRs, covering first-order to fourth-order MMRs—and present their resonant angle distributions in Fig.~\ref{fig:type_phi}. As expected, the blue points representing $\phi_0$ resonators occupy the largest region in the $e-i$ plane, while the red points representing $\phi_1$ resonant objects generally appear only in regions of high-$i$ and low-$e$, and the green points representing $\phi_{multi}$ resonators are scattered in between. For most MMRs and most $\dot{a}_N$, $\phi_1$ resonators usually appear only in regions with $i \gtrsim 20^\circ$. However, for different MMRs, the distribution of resonant angles still shows some differences. 

\subsection{First-order resonances}

For a $p$:$p+1$ MMR, the lowest-order resonant angle is only $\phi_0$. However, higher-order MMRs may also exist at the same location, such as $2p$:$2p+2$ MMR; for this second-order MMR, the inclination-type resonant angle $\phi_1 = (2p+2)\lambda - 2p\lambda_N - 2\Omega$ does exist. For example, for the 1:2 MMR, the inclination-type MMR of the 2:4 MMR may be present, with the corresponding resonant angle $\phi_1 = 4\lambda - 2\lambda_N - 2\Omega$. Owing to the higher $k$, this resonant angle can only exist when the object has a very low $e$ and a high $i$. \citetads{Li2024} have shown that this resonant angle may librate when objects switch between asymmetric islands of 1:2 MMR through a low-$e$ channel. Therefore, for first-order MMRs, we also check whether the $\phi_1$ of the corresponding second-order MMR is librating.

As shown in Fig.~\ref{fig:type_phi}, the number of $\phi_1$ resonators is very small for a first-order MMR, as a result of their higher order. These $\phi_1$ resonators are confined to a narrow region of low $e$ and high $i$. Around these $\phi_1$ resonators, some $\phi_0$ resonators and $\phi_{multi}$ resonators are also distributed.

For a $\phi_{multi}$ resonator, a particularly meaningful question is which resonant angle is dominant. When $\dot{a}_N \leq 1$\,au/Myr, the distribution of $\phi_{multi}$ resonators extends all the way to the lower-left corner of the $e-i$ plane, where $i\approx0^\circ$. However, it is clear that in this region, resonance capture is still dominated by eccentricity, so these $\phi_{multi}$ resonators should have been captured by the eccentricity-type resonance first; because $\dot{\omega}$ is very small, the $\phi_1$ merely librates concomitantly, or circulates at an extremely slow rate. In the vicinity of where the $\phi_1$ resonant objects are distributed, since several resonance types are mixed together, it is not straightforward to directly determine which type is dominant. For a $\phi_{multi}$ resonator, one possible criterion might be whether $\Delta\phi_0$ or $\Delta\phi_1$ is smaller.

When $\dot{a}_N = 10$\,au/Myr, the capture region for the $\phi_0$ resonators shifts toward higher $e$. Here we can observe that the low-$e$ region is composed of $\phi_{multi}$ resonators and $\phi_1$ resonators, but the capture probability is very low. If we examine the libration amplitudes of these $\phi_{multi}$ resonators, we find that $\Delta\phi_0$ is generally slightly smaller than $\Delta\phi_1$, reflecting that this region is still dominated by inclination-type MMR even though $\phi_1$ resonators are extremely rare.

\subsection{Second-order resonances}

Starting from the second-order MMRs, the $\phi_1$ resonators become more common. In the slow migration cases ($\dot{a}_N \leq 0.1$\,au/Myr), we find that in the low-$e$ region, $\phi_{multi}$ resonators begin to appear when $i\gtrsim 16^\circ$, and $\phi_1$ resonators become widespread when $i \gtrsim 28^\circ$, and both populations continue extending toward higher-$i$ regions. Their distribution range overlaps with that of $\phi_0$ objects.

Comparing the 3:5 MMR of Fig.~\ref{fig:type_phi} with Fig.~\ref{fig:ei_dis_2nd}, we find that the distribution region of $\phi_1$ resonators is almost identical to the region where the capture probability is highest. It lies at the very left edge of the region allowed for capture. This suggests that the capture by inclination-type and eccentricity-type resonances may not be merely a competition; rather, at the microscopic level, when capture occurs, each type may favor different subregions of phase space. The two types of capture superimpose in this ``band'' region, producing the relatively high capture probability observed there.

In the fast migration cases ($\dot{a}_N \geq 1$\,au/Myr), the distribution regions of $\phi_0$ resonant objects and $\phi_1$ resonant objects begin to separate. At this stage, eccentricity-type resonance capture occurs in higher-$e$ regions, while inclination-type resonance capture favors lower-$e$ regions. This is similar to the phenomenon observed for the 1:2 MMR at $\dot{a}_N = 10$\,au/Myr.

\subsection{Third-order resonances}

In the previous section we have already noted that, the $e-i$ plane can be divided into $e$-dominated and $i$-dominated regions when $k\geq3$. A straightforward conjecture is that, in the $e$-dominated region, the dominant resonant angle should be $\phi_0$, while in the $i$-dominated region it should be $\phi_1$ or $\phi_2$. In Fig.~\ref{fig:type_phi}, the region to the left of the dotted red line is dominated by $\phi_1$ resonators and $\phi_{multi}$ resonators, reflecting a strong correlation between the $i$-dominated region and the mixed-type resonant angle $\phi_1$. However, this strong correlation does not imply complete determinism: there are still some $\phi_1$ resonators within the $e$-dominated region, and likewise some $\phi_0$ resonators within the $i$-dominated region. As the lowest-order MMRs exhibiting the differentiation phenomenon, the differentiation of capture regions is most pronounced for the third-order MMRs. In the slow migration cases ($\dot{a}_N \leq 0.1$\,au/Myr), the $i$-dominated region also contains an inevitable capture zone, which produces the largest number of $\phi_1$ resonators.

Within the $i$-dominated region, $\phi_1$ resonators become widespread when $i \gtrsim 24^\circ$, while $\phi_{multi}$ resonators cover a broad range of inclinations. By examining $\Delta\phi_0$ and $\Delta\phi_1$ of these $\phi_{multi}$ resonators, we confirm that they are still predominantly governed by the mixed-type resonance. Nevertheless, in the lowest-$i$ region, there are still some particles with $\Delta\phi_0 < \Delta\phi_1$.

As the $\dot{a}_N$ increases, the inevitable capture zone is the first to disappear ($\dot{a}_N = 1$\,au/Myr). When $\dot{a}_N = 10$\,au/Myr, the $i$-dominated region also vanishes completely. At this time, the third-order MMRs are similar to the first- and second-order MMRs, with $\phi_{multi}$ resonators occupying the lower-$e$ region. It is worth noting that, for the 4:7 MMR, not a single $\phi_1$ resonant object is recorded at $\dot{a}_N = 10$\,au/Myr; however, by examining the libration amplitudes of the $\phi_{multi}$ resonators, we find that the leftmost objects are still dominated by the mixed-type resonant angle $\phi_1$.

\subsection{Fourth-order resonances}

When the order reaches 4, there are three possible resonant angles: $\phi_0$, $\phi_1$, and $\phi_2$. In the slow migration cases ($\dot{a}_N \leq 0.1$\,au/Myr), resonant bodies corresponding to all three resonant angles exist, with $\phi_0$ resonators located in the lower right, $\phi_2$ objects in the upper left, and $\phi_1$ resonators situated in between, illustrating a transition among resonant angles from eccentricity-type to mixed-type and finally to inclination-type.

Similar to the 4:7 MMR, the $e-i$ plane of the 3:7 MMR can be roughly divided into $e$-dominated and $i$-dominated regions. The $\phi_2$ objects are found only in the $i$-dominated region, close to its left edge. This indicates that the existence of pure inclination-type resonance is in fact very restrictive, requiring very low $e$ and high $i$. Meanwhile, $\phi_1$ resonators are distributed in both the $e$-dominated and $i$-dominated regions, and a small number of $\phi_0$ resonators are also found within the $i$-dominated region, similar to the results in the third-order MMRs. These phenomena again demonstrate that the form of the resonant angle for an object is not entirely determined by whether its capture region is $i$-dominated or $e$-dominated.

In the fast migration cases ($\dot{a}_N \geq 1$\,au/Myr), the 3:7 MMR does not differ greatly from the lower-order MMRs; however, the transition from slow to fast migration becomes less abrupt, which is a characteristic feature of high-order MMRs. At $\dot{a}_N = 1$\,au/Myr, $\phi_2$ objects disappear first, and by $\dot{a}_N = 10$\,au/Myr, $\phi_1$ resonators also vanish. Overall, regardless of the resonance order, eccentricity-type resonances are more capable of capturing objects under fast migration.

\section{Capture probability and efficiency}

We have now reached a series of conclusions regarding resonance capture, involving different MMRs, different $\dot{a}_N$, different $e$ and $i$, and different resonance types. However, most of the above conclusions are based on a uniform $e-i$ distribution and artificial ranges of $e$ and $i$, which limits any quantitative comparison across different MMRs. For example, capture into the 3:4 MMR is confined to a very narrow eccentricity interval, yet its capture probability is relatively high; by contrast, capture into the 1:5 MMR can occur over a broader region but with a lower probability. In addition, the likelihood that an MMR captures a $\phi_1$ or $\phi_{multi}$ resonator is also closely related to the excitation level of the objects. Fig.~\ref{fig:type_phi} has already shown that the wider the inclination range, the more inclination-type or mixed-type captures occur, and in a planar model, naturally all particles end up being captured into purely eccentricity-type resonances.

To turn the above qualitative trends into quantitative statements, it is necessary to place all MMRs within a unified model of the small body distribution. To this end, we assume that both $e$ and $i$ follow Rayleigh distributions and are independent of each other. We test three different levels of planetesimal excitation, corresponding to a ``cold'', ``moderate'', and ``hot'' disk, with Rayleigh distribution parameters for eccentricity, $\sigma_e$, of 0.04, 0.1, and 0.2, corresponding to mean value of $e$ of about 0.05, 0.13, and 0.25, respectively. The Rayleigh parameter for inclination is taken as $\sigma_i = \frac{1}{2}\sigma_e$, yielding mean value of $i$ of about $1.4^\circ$, $3.6^\circ$, and $7.2^\circ$, respectively. This empirical assumption implies that the mechanisms for exciting inclination and eccentricity are the same.

\begin{figure}[!htb]
\centering
\resizebox{\hsize}{!}{\includegraphics{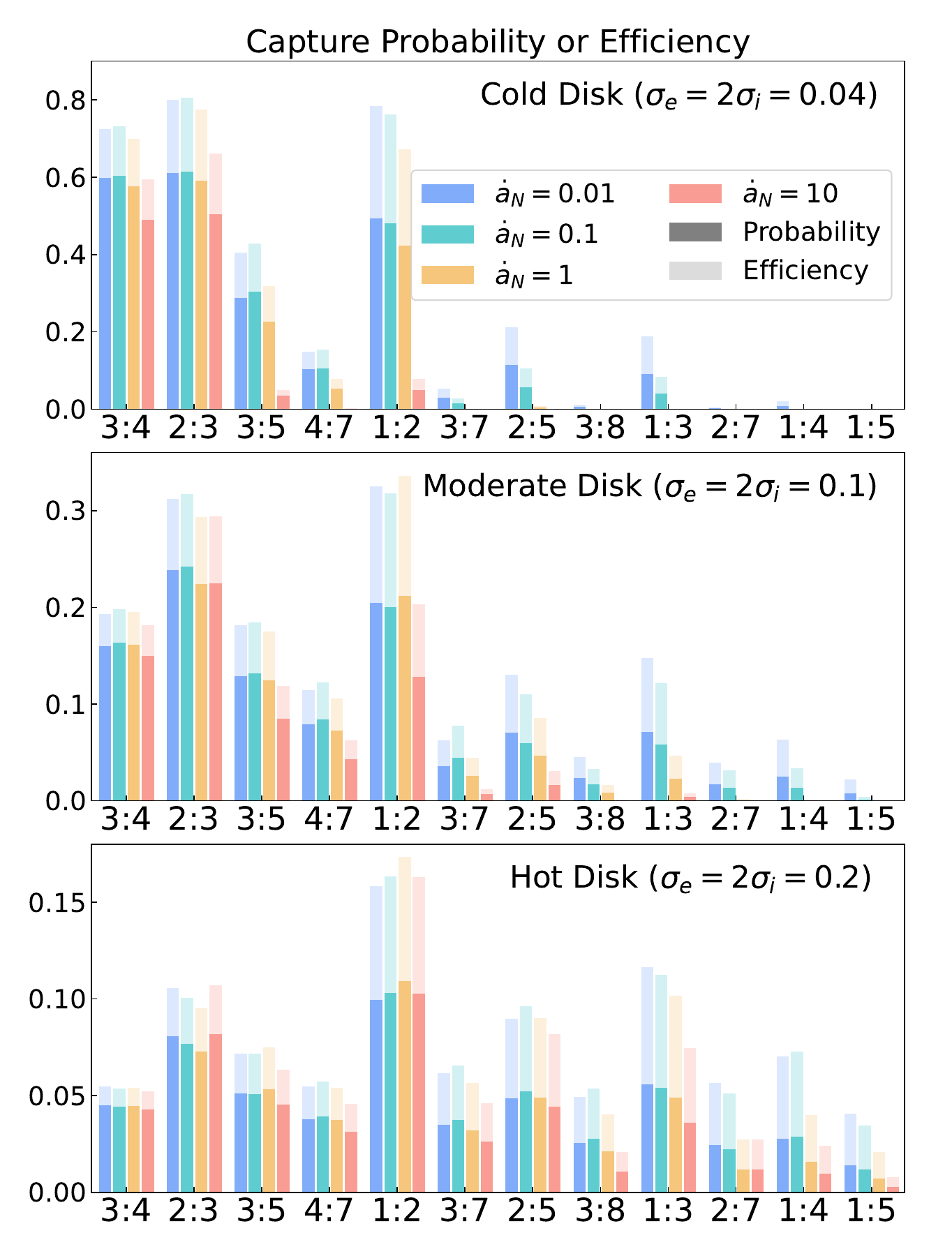}}
\caption{Capture probabilities and efficiencies of different MMRs. The horizontal axis indicates the different MMRs, arranged at equal intervals in order of increasing distance, and the vertical axis shows the capture probability or efficiency. The dark-colored bars represent the capture probability of an MMR, and the light-colored bars represent the capture efficiency. For each MMR, the four differently colored bars indicate the different migration rates, increasing from left to right. The three subplots represent the different $e-i$ distributions (from top to bottom, the excitation level increases).}
\label{fig:cap_prob}
\end{figure}

In section 3, we have obtained the probability density distribution in the $e$–$i$ plane (as in Fig.~\ref{fig:ei_dis_1st}, \ref{fig:ei_dis_2nd}, \ref{fig:ei_dis_3rd}, and \ref{fig:ei_dis_high}). Multiplying the probability density distribution by the capture probability distribution, and integrating over the plane yields the expectation that a particle will be captured into a given MMR. A variable following a Rayleigh distribution has no upper bound; we therefore adopt the same upper integration limits as in the studied parameter space, i.e., $e$ from 0 to $e_{max}$ (given in Table \ref{tab:MMR_ini}) and $i$ from $0^\circ$ to $60^\circ$. We have previously noted that at high $i$, the eccentricity range over which an MMR can capture small bodies may exceed $e_{max}$, and this portion of the probability is discarded in the calculation. Nevertheless, under a Rayleigh distribution, the probability of a small body with both high-$i$ and high-$e$ is extremely low, so this truncation does not appreciably affect the results. In Fig.~\ref{fig:cap_prob} we present the capture probabilities of the MMRs under different $\dot{a}_N$ and excitation levels of the planetesimal disk.

In the cold disk, a prominent feature is that the capture probability is sensitive to both the specific MMR and $\dot{a}_N$. The two closest first-order MMRs both have very high capture probabilities, generally around one half, as a result of the inevitable capture at low-$e$. The distant 1:2 MMR also has a capture probability close to one half when $\dot{a}_N \leq 1$\,au/Myr, but its capture probability drops sharply to only about 5\% at $\dot{a}_N = 10$\,au/Myr. Correspondingly, the capture probabilities of the second- and third-order MMRs decrease as well, and they all undergo a sharp drop at $\dot{a}_N = 1$ or 10\,au/Myr. On the other hand, MMRs with $k\geq4$ are essentially unable to capture particles, with their capture probabilities all close to zero, because in this cold disk, most objects do not meet the $e_{min}$ required for capture.

In the moderate disk, the particles become more dispersed into a somewhat larger-$e$, causing the capture numbers of the two closest first-order MMRs to decrease across all values of $\dot{a}_N$. The 1:2 MMR is slightly different, because its capture capability at $\dot{a}_N = 10$\,au/Myr is better than in the cold disk. A similar pattern holds for several second- and third-order MMRs: compared with the cold disk, their capture probabilities in the moderate disk increase at faster $\dot{a}_N$ and decrease at slower $\dot{a}_N$, which reflects the selective effect of $\dot{a}_N$ with respect to eccentricity range. For the weaker fourth- and fifth-order MMRs, regardless of the $\dot{a}_N$, their capture performance is better than in the cold disk.

In the hot disk, the capture probabilities of the first-order MMRs all decline more or less. Among them, the capture probability of the 3:4 MMR drops to below 5\%, while the capture probability of the 1:2 MMR declines the slowest and has already surpassed that of the 2:3 MMR, as a result of the higher eccentricity of the particles. In contrast, the capture probabilities of the fourth- and fifth-order MMRs increase to different extents. Overall, the differences in capture probability among different MMRs become relatively small. Meanwhile, for most MMRs, the capture probability is less sensitive to the $\dot{a}_N$, with only a few high-order MMRs showing a clear decline in capture probability as $\dot{a}_N$ becomes faster.

Mean-motion resonances at different locations naturally sweep across different radial ranges. To compare the capture capabilities of different MMRs more directly, we define the ``capture efficiency'' of an MMR as the product of its capture probability and its $\alpha$ value, as shown by the light-colored bars in Fig.~\ref{fig:cap_prob}. It should be noted that a more reasonable way to compare the capture capabilities should take into account that the planetesimal density in the region swept by each MMR may differ; however, the exact distribution of planetesimal density is unknown. Therefore, the approach adopted here—simply multiplying the capture probability by the $\alpha$ value—effectively assumes that the radial density of the planetesimals is uniform, which also means that the surface density is inversely proportional to the heliocentric distance. Moreover, we have not considered that the excitation level of disk may also vary at different distances.

Compared with the capture probability, the capture efficiency correspondingly improves the estimated capture capability for the exterior MMRs. However, compared with the factors mentioned earlier—such as $e$, $i$, and $\dot{a}_N$—there is no order-of-magnitude difference between these two metrics. It can be noted that, in the cold and moderate disks, except for the case of $\dot{a}_N = 10$\,au/Myr, the capture efficiency of the 1:2 MMR is comparable to that of the 2:3 MMR. In the hot disk, however, the capture efficiency of the 1:2 MMR far exceeds that of the 2:3 MMR. Considering that the observational estimate gives a population for the 1:2 MMR that is only about half that of the 2:3 MMR, this discrepancy in fact reflects the difference in the stability of these two major MMRs \citepads[e.g.,][]{Melita2000}.

\begin{figure}[!htb]
\centering
\resizebox{\hsize}{!}{\includegraphics{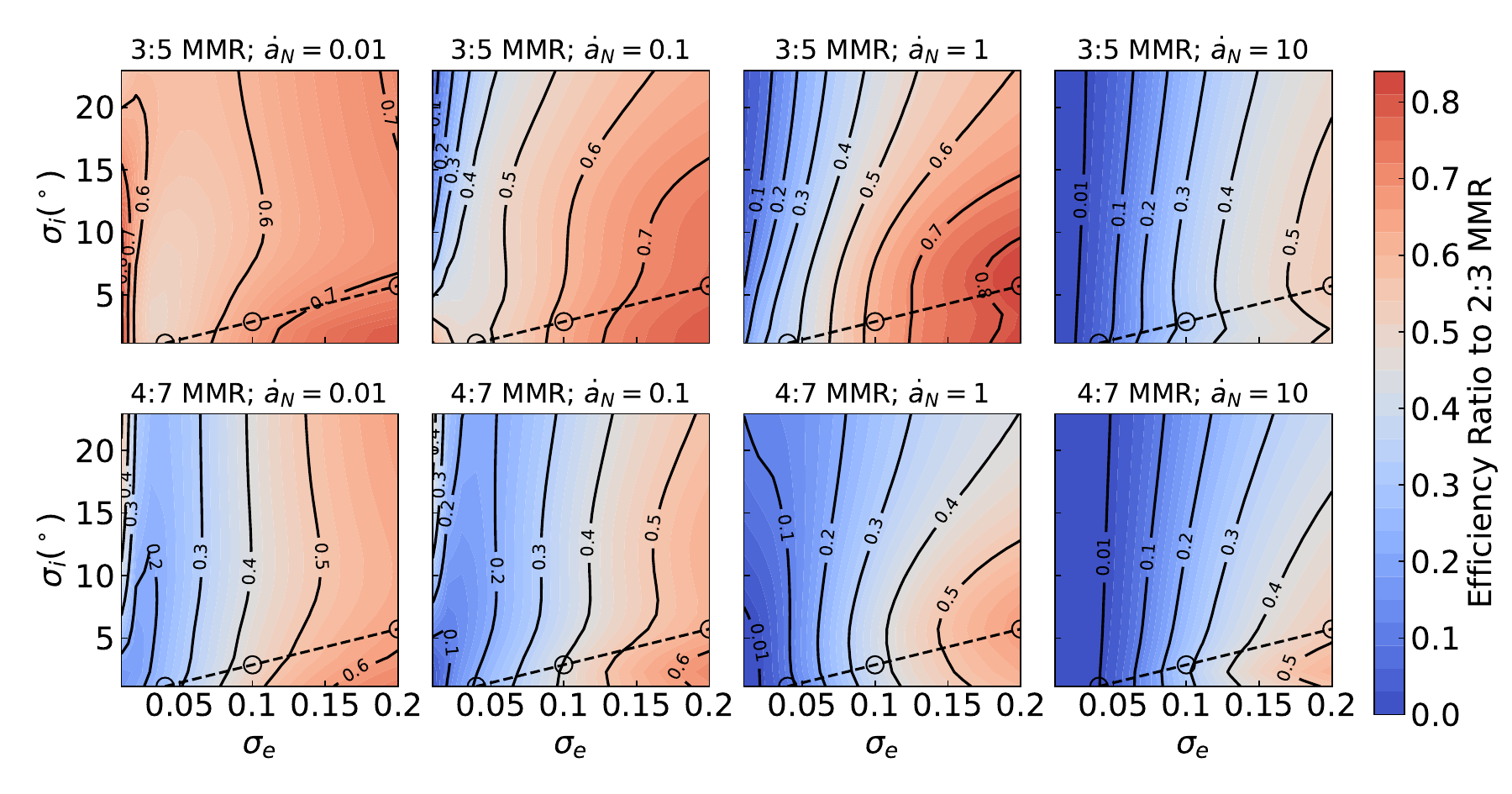}}
\caption{Capture efficiency ratios of the 3:5 MMR (upper panels) and 4:7 MMR (lower panels) relative to the 2:3 MMR. The horizontal and vertical axes represent the Rayleigh distribution parameters $\sigma_e$ and $\sigma_i$ for $e$ and $i$, respectively. The four columns correspond to the four different $\dot{a}_N$. The dashed line marks the location where $\sigma_i = \frac{1}{2}\sigma_e$, and the circular markers denote the parameter combinations used in Fig.~\ref{fig:cap_prob}.}
\label{fig:cap_eff_r}
\end{figure}

In Section 3, we have shown that an increasing $i$ generally reduces the capture probability of an MMR. However, the degree of reduction varies among MMRs, and therefore affects the population ratios among MMRs. Next, we attempt to separate the effects of $e$ and $i$ and compare how they impact different MMRs. To this end, we select three relatively nearby MMRs—the 2:3, 3:5, and 4:7 MMRs—for comparison, and discuss different $e$ and $i$ excitation. Specifically, we took 20 values of $\sigma_e$ from 0.01 to 0.20 and 20 values of $\sigma_i$ from 0.02 to 0.4. For each parameter combination, we use the 2:3 MMR as a reference and calculate the capture efficiency ratios of the 3:5 and 4:7 MMRs relative to it; the resulting maps are shown in Fig.~\ref{fig:cap_eff_r}.

Compared with the 2:3 MMR, the 3:5 and 4:7 MMRs are weaker MMRs. Fig.~\ref{fig:cap_eff_r} shows that the factor primarily affecting their capture efficiency ratios relative to the 2:3 MMR is $\sigma_e$. As for inclination, although we deliberately expanded the range of $\sigma_i$ in the calculation, its influence is evidently rather weak. The effect of $e$ is significantly stronger than that of $i$, causing the contours in Fig.~\ref{fig:cap_eff_r} to be oriented roughly vertically. Thus, the result shown in Fig.~\ref{fig:cap_prob}—that a more excited disk favors capture into higher-order MMRs—is essentially due to its higher $\sigma_e$.

So far, we have shown that, in general, a larger $\sigma_e$ contributes to the capture of weak MMRs. Nevertheless, there are still some details worth noting in Fig.~\ref{fig:cap_eff_r}. For example, when $\dot{a}_N \leq 0.1$\,au/Myr and $\sigma_e\sim0$ with $\sigma_i$ relatively small, the capture efficiency of the 3:5 MMR relative to the 2:3 MMR is enhanced to some extent. This in fact reflects that, on the one hand, when particles are widely distributed within the inevitable capture zone, the capture efficiency of the 2:3 MMR has already reached its maximum, while on the other hand, the capture probability of the 3:5 MMR improves markedly at low-$e$ and low-$i$, as shown in Fig.~\ref{fig:ei_dis_2nd}. As a result, the efficiency ratio between the 3:5 MMR and the 2:3 MMR improves, leading to a phenomenon that contradicts the general trend.

As the excitation level of inclination increases, the capture efficiencies of all MMRs decline. For strong MMRs such as the 2:3 MMR, higher-$i$ narrows its inevitable capture zone, thereby reducing the capture efficiency, whereas for weaker MMRs such as the 3:5 MMR, higher-$i$ raises the $e_{min}$. Comparing the two mechanisms, the weaker MMRs are more strongly affected, which explains the slight decline in the efficiency ratio between the 3:5 MMR and the 2:3 MMR at higher $\sigma_i$ as shown in Fig.~\ref{fig:cap_eff_r}. Consequently, most contours, while maintaining a generally vertical orientation, also tilt slightly clockwise. This also indicates that, when the effects of $e$ and $i$ are considered separately, their overall effects are different. A larger $i$ leads to a greater disparity in capture efficiency between strong and weak MMRs.

However, as the lowest-order MMRs in which capture region differentiation can occur, the third-order MMRs possess a robust inevitable capture zone at high-$i$ and low-$e$ region, which also makes them the MMRs most likely to benefit from an increase in $\sigma_i$. Fig.~\ref{fig:cap_eff_r} shows that, while the contours also exhibit a clockwise tilt, the degree of tilting for the 4:7 MMR is smaller. Moreover, when $\dot{a}_N \leq 1$\,au/Myr and $\sigma_e$ is small, the efficiency ratio of the 4:7 MMR to the 2:3 MMR increases with $\sigma_i$. These phenomena reflect that, relatively speaking, the effect of increased $\sigma_i$ on third-order MMRs is minimal; and under certain specific conditions the increase in $\sigma_i$ may even help enhance their capture efficiencies.

This section demonstrates that the planetesimal distribution significantly affects the capture efficiency ratios among different MMRs. A cold disk favors the capture of strong MMRs, while a disk with higher excitation levels tends to balance the disparities between MMRs and makes the capture efficiency less sensitive to $\dot{a}_N$. When $e$ and $i$ are considered separately, the above effect is dominated by eccentricity. The effect of inclination is somewhat different: a higher $i$ generally reduces the capture efficiency of MMRs, but the weaker MMRs are more strongly affected, so higher $i$ is relatively more favorable for the strong MMRs. Additionally, the third-order MMRs, owing to their high capture probability within the $i$-dominated region, are most likely to benefit from increasing $i$. Overall, the excitation level of the disk is an important factor affecting the population of different MMRs. It also implies that we may be able to probe the excitation level in the primordial disk by studying the population ratios across different MMRs.

\section{Conclusion and discussion}

In this paper, we aim to build upon the work of \citetads{Li2026} by extending the model to three dimensions and investigating the effects of $e$ and $i$ on resonance capture. Through a series of numerical simulations under the three-body model, we present the capture probability distributions for 12 MMRs under the effects of different $\dot{a}_N$, $e$ and $i$ (Fig.~\ref{fig:ei_dis_1st}, \ref{fig:ei_dis_2nd}, \ref{fig:ei_dis_3rd}, and \ref{fig:ei_dis_high}). We studied how the capture probability distribution is affected by the intrinsic properties of the MMR (such as $k$ and $\alpha$), by the particle distribution (such as their $e$ and $i$), and by the $\dot{a}_N$.

Previous work has shown that Neptune generally follows an exponentially decaying migration \citepads[e.g.,][]{Malhotra1993}, while stochastic processes further render the migration grainy \citepads[e.g.,][]{Zhou2002, Nesvorny2016, Lawler2019}, and there may also be a major jump caused by giant planet instability \citepads[e.g.,][]{Tsiganis2005, Nesvorny2012, Nesvorny2015a}. For this study, however, we only tested the simplest case of uniform $\dot{a}_N$ over a short migration distance, in order to present the most direct consequences of the $\dot{a}_N$. Since the $\dot{a}_N$ we tested cover a wide range (from 0.01\,au/Myr to 10\,au/Myr), a realistic migration model can be obtained by connecting simple models with different $\dot{a}_N$. Reconstructing this realistic migration process is also a very important direction for future efforts.

Under the $\dot{a}_N$ values we tested, inevitable capture will happen for relatively strong MMRs (such as the 3:4 and 2:3 MMRs) within a certain eccentricity range, and this eccentricity threshold narrows as the inclination increases. Meanwhile, relatively weak MMRs (such as the 1:3 and 3:7 MMRs) require $e\geq e_{min}$ so that capture can occur, and $e_{min}$ increases with inclination. There are also MMRs with moderate resonance strength (such as the 1:2 and 3:5 MMRs), which behave more like a strong MMR at slow $\dot{a}_N$, while resembling a weaker MMR at fast $\dot{a}_N$. Taken together, for the vast majority of MMRs, an increase in inclination raises both eccentricity thresholds discussed above, thereby hindering resonance capture.

At the high-$e$ end, on the other hand, the main obstacle to resonance capture is the scattering effect, because an excessively high $e$ causes the particles to be scattered first, thereby losing the opportunity to be captured. In the planar case, the maximum eccentricity threshold allowing capture is roughly the critical eccentricity at which the small body's perihelion approaches 2\,au beyond Neptune's orbit \citepads{Li2026}. When the small body possesses a certain inclination, its probability of being scattered tends to decrease, thus raising this threshold. This means that at the high-$e$ end, an increasing $i$ is possibly favorable for resonance capture.

A faster migration likewise narrows the inevitable capture zones of strong MMRs, raises the $e_{min}$ required for weak MMRs, and can cause some strong MMRs to gradually exhibit characteristics of weak MMRs. Overall, faster migration also hinders resonance capture. One can further imagine that at even more extreme values of $\dot{a}_N$, the inevitable capture zones of the 2:3 and 3:4 MMRs would eventually disappear, and these MMRs would also require a certain $e_{min}$ to capture objects. Conversely, under extremely slow $\dot{a}_N$, it can be expected that the $e_{min}$ will drop to 0, or even develop inevitable capture zones for those relatively weak MMRs.

Nevertheless, the $\dot{a}_N$ range tested in this work already covers the $\dot{a}_N$ that Neptune could plausibly have experienced. Previous evidence from the distribution of TNOs mostly indicates that Neptune's migration cannot be particularly fast \citepads[e.g.,][]{Nesvorny2015b, Lawler2019}, and simulations of planetary migration also suggest that driving extremely fast migration requires an exceptionally fast consumption of planetesimals \citepads{Li2025}. On the other hand, during specific phases of Neptune’s migration, $\dot{a}_N$ is expected to drop below the slowest value considered here (0.01\,au/Myr). Nevertheless, according to \citetads{Li2025}, planetary migration is a positive-feedback process: once $\dot{a}_N$ decreases to a very low value, Neptune clears the nearby orbits within a relatively short time and reduces the probability of further close encounters with planetesimals, thereby suppressing migration. This implies that even if $\dot{a}_N$ could drop below 0.01\,au/Myr, such migration could not be sustained over a long distance. Accordingly, Neptunian MMRs would not sweep across a large radial range at such a slow rate, so the associated capture can likely be neglected.

When $k$ reaches three, the capture probability distribution in the $e-i$ plane undergoes differentiation: the high-$e$ or low-$i$ region corresponds to the $e$-dominated region, whereas in the low-$e$ and high-$i$ region, capture is $i$-dominated. Between these two regions, there is a low capture probability gap that separates them. Moreover, the $i$-dominated region possesses a considerable capture probability and is able to inevitably capture particles within certain $e$ and $i$ values (Fig.~\ref{fig:boundfind}). As the $\dot{a}_N$ increases, the inevitable capture region disappears first, after which the entire $i$-dominated region vanishes completely. When $k$ reaches five, further differentiation occurs in the $e-i$ plane, and we speculate that each mixed-type resonant angle may correspond to a newly emerging capture region.

Subsequently, we carried out an examination of the resonant angle types (Fig.~\ref{fig:type_phi}). We find that for 1:2 MMR, sporadic $\phi_1$ resonators already exist, corresponding to the inclination-type resonant angle of the 2:4 MMR. For second-order MMRs, $\phi_1$ resonators become more common; they are located in the low-$e$, high-$i$ corner of the capture region. For third-order MMRs, as the capture region undergoes differentiation, the resonant angle of a small body is highly correlated with the region in which it is captured: a large number of $\phi_1$ resonators are found in the $i$-dominated region, while the $e$-dominated region is almost entirely populated by $\phi_0$ resonators. However, this relationship is not absolute, and in the actual Solar System, due to perturbations from other giant planets, $\phi_0$ and $\phi_1$ resonators may also transform into each other. For MMRs with $k\geq4$, one can also see a transition among multiple $\phi_j$ types: from the lower-right to the upper-left corner, there are successively $\phi_0$, $\phi_1$ and $\phi_2$ resonators. In addition, there are a considerable number of $\phi_{multi}$ resonators, whose multiple resonant angles librate simultaneously, typically filling the transition regions between the different $\phi_j$ resonators.

During resonance capture, the superposition of different resonance types further complicates the situation. There are at least two potential superpositions in this paper: first, the superposition of MMRs of different orders, for example, the superposition between the 1:2 MMR and the 2:4 MMR; and second, the superposition of different resonant angles (eccentricity-type, mixed-type, and inclination-type), as suggested by Fig.~\ref{fig:type_phi}. We note that in the $e-i$ plane, whenever such overlap occurs in a certain region, the capture probability is typically higher than that of the surrounding regions. This suggests that the different types of capture may not be in a purely competitive relationship, but rather they may occur independently, so that their combination can enhance the capture probability. Moreover, for the 1:$q$ MMRs, there may also be a superposition phenomenon involving different resonance centers \citepads[e.g.,][]{Li2023}. In a complete Solar System model, besides two-body MMRs with Neptune, objects may also reside in two-body MMRs with Uranus, as well as in three-body MMRs involving both Uranus and Neptune. The superposition of these MMRs makes the motion of the TNOs more complex \citepads[e.g.,][]{Smirnov2025}.

The above results emphasize that the distribution of planetesimals is crucial for capture. For typical MMRs, the capture probability can be either extremely high or extremely low; for MMRs with $k\geq2$, the fraction of $\phi_1$ resonators can also be very large or very small. Therefore, regardless of which aspect of resonance capture we discuss, the distribution of the small bodies cannot be excluded from consideration. To this end, we assume that both $e$ and $i$ follow Rayleigh distributions with $\sigma_i = \frac{1}{2}\sigma_e$, and we further tested three different excitation levels, corresponding to a cold disk ($\sigma_e=0.04$), a moderate disk ($\sigma_e=0.1$), and a hot disk ($\sigma_e=0.2$), respectively. Based on these distributions, we calculated the capture probabilities for all MMRs accordingly (Fig.~\ref{fig:cap_prob}). We find that a cold disk strongly favors the capture of closer, low-order MMRs, whereas a hot disk balances the differences among the MMRs.

Recently, \citetads{Li2026} carried out a similar comparative study. They likewise conclude that a higher excitation level of eccentricity facilitates the capture of exterior weak MMRs. However, in their work, this conclusion was obtained with a planar model: first, they used a fitting method to determine the eccentricity range over which an MMR can capture particles; and then they assumed a uniform capture probability within that range. Although our present work focuses on only a subset of the MMRs, we have extended the study to three dimensions and more carefully accounted for how the capture probability varies with $e$ and $i$. The consistency of the above results further supports the agreement between the two estimation approaches.

By multiplying the capture probability of an MMR by its $\alpha$, we obtained its capture efficiency, a metric that allows for a more direct comparison of the captured population of different MMRs (Fig.~\ref{fig:cap_eff_r}). By testing different $\sigma_e$ and $\sigma_i$, we find that the main factor controlling the capture ratios among different MMRs is eccentricity rather than inclination. Moreover, generally speaking, a higher $\sigma_e$ tends to balance the capture ratios between strong and weak MMRs, whereas a higher $\sigma_i$ tends to widen the gap between them. However, third-order MMRs benefit most from an inclination increase, as a result of the $i$-dominated region and their inevitable capture zone.

Direct comparisons with observations often rely on a series of idealized assumptions. When we calculated capture efficiency, we assumed that the planetesimals in the primordial disk possess a uniform semimajor distribution and a consistent $e-i$ distribution. In reality, the disk cannot extend outward infinitely, and the manner in which the density decays with distance would also significantly affect the capture number of different MMRs. Moreover, different regions of the disk may have different $e-i$ distributions; at locations close to Neptune, the scattering effect of the planet can substantially excite the eccentricities of the planetesimals. Conversely, when an MMR sweeps through particles without capturing them, it may decrease their eccentricities \citepads[][]{Murray1999,Mustill2011}. Furthermore, stellar flybys may lead to higher excitation levels for more distant planetesimals.

Another issue worth noting is the probability that an object can be captured as a $\phi_1$ resonator. If we assume that the excitation of $e$ and $i$ arises from the same dynamical mechanism, we would obtain $\sigma_i = \frac{1}{2}\sigma_e$. Then when the mean eccentricity reaches 0.25, the mean inclination is only $7.2^\circ$. However, Fig.~\ref{fig:type_phi} shows that the common occurrence of $\phi_1$-type capture requires $i\gtrsim20^\circ$. In a relatively planar disk, $\phi_1$ resonators are unlikely to appear, whereas the existence of $\phi_{multi}$ resonators may be more probable.

Furthermore, when comparing the capture capabilities of different order MMRs for $\phi_1$ resonators, it is evident that third-order MMRs are the most likely to host $\phi_1$ resonators, because they are the lowest-order MMRs in which the differentiation of capture regions occurs. Although MMRs with $k\geq4$ also exhibit this differentiation, their $i$-dominated regions require more stringent conditions and are almost nonexistent within typical ranges of $\dot{a}_N$ and inclination. In contrast, the $i$-dominated region of the 4:7 MMR exists over a wider range; at $\dot{a}_N \leq 0.1$\,au/Myr, the 4:7 MMR can inevitably capture objects with $i \gtrsim 5^\circ \sim 10^\circ$. Thus, among all currently observed TNOs, 2014 TZ85 is the only $\phi_1$ resonator (its eccentricity-type angle is circulating), and it is located in the 4:7 MMR \citepads{Li2020}, which is not a coincidence. We can reasonably anticipate that the 2:5 and 1:4 MMRs may also host $\phi_1$ resonators, whereas the probability of finding $\phi_1$ resonators in second-order MMRs (such as the 3:5 and 1:3 MMRs) is correspondingly lower.

This study highlights the influence of particles' eccentricity and inclination on resonance capture. In the future, as the observational sample grows, it will help the population ratios to be more accurately determined among different MMRs, which in turn will allow us to probe the excitation level of planetesimal disk in the early Solar System. Moreover, the type of resonance capture is also strongly correlated with the orbital distribution of the particles. Thus, in discovering more $\phi_1$ resonant bodies, understanding their fraction and distribution will greatly help us constrain the inclination distribution within the primordial disk.

\begin{acknowledgements}
        We gratefully acknowledge Dr. Ivana Milić Žitnik for her careful review of the manuscript and helpful suggestions, which have enhanced the scientific rigor of this paper. This work is supported by the Science and Technology Development Fund (FDCT) of Macau (grant Nos. 0034/2024/AMJ, 0008/2024/AKP, 002/2024/SKL, 0002/2025/AKP) and the National Key Research and Development Program of China (grant No.2024YFE0201000). Li-Yong Zhou acknowledges the support from the National Natural Science Foundation of China (NSFC, Grants No.12373081 \& No.12150009) and the China Manned Space Program with grant No. CMS-CSST-2025-A16.
\end{acknowledgements}

\bibliographystyle{aa-note}
\bibliography{ResCap2}
\end{document}